\documentclass[10pt,prd,aps,twocolumn,showpacs,nofootinbib,superscriptaddress,floatfix,nopreprintnumbers]{revtex4}

\usepackage{graphicx}
\usepackage{bm} 
\usepackage{subfigure}

\makeatletter
\newcommand\erfc{\mathop{\operator@font erfc}\nolimits}
\def\slashchar#1{\setbox0=\hbox{$#1$}
   \dimen0=\wd0 \setbox1=\hbox{/} \dimen1=\wd1
   \ifdim\dimen0>\dimen1 \rlap{\hbox to \dimen0{\hfil/\hfil}} #1
   \else  \rlap{\hbox to \dimen1{\hfil$#1$\hfil}} / \fi}

\newcommand{\taug}{\tau_{g,\rm eq}}
\newcommand{\tauq}{\tau_{q,\rm eq}}

\begin{document}
 
\title{Hydrodynamics of anisotropic quark and gluon fluids}

\author{Wojciech Florkowski} 
\affiliation{Institute of Physics, Jan Kochanowski University, PL-25406~Kielce, Poland} 
\affiliation{The H. Niewodnicza\'nski Institute of Nuclear Physics, Polish Academy of Sciences, PL-31342 Krak\'ow, Poland}

\author{Radoslaw Maj} 
\affiliation{Institute of Physics, Jan Kochanowski University, PL-25406~Kielce, Poland} 

\author{Radoslaw Ryblewski} 
\affiliation{The H. Niewodnicza\'nski Institute of Nuclear Physics, Polish Academy of Sciences, PL-31342 Krak\'ow, Poland} 

\author{Michael Strickland} 
\affiliation{Physics Department, Kent State University, Kent, OH 44242 United States}
\affiliation{Frankfurt Institute for Advanced Studies, Ruth-Moufang-Strasse 1, D-60438, Frankfurt am Main, Germany}

\date{\today}

\begin{abstract}
The recently developed framework of anisotropic hydrodynamics is generalized to describe the dynamics of coupled quark and gluon fluids. The quark and gluon components of the fluids are characterized by different dynamical anisotropy parameters. The dynamical equations describing such mixtures are derived from kinetic theory with the collisional kernel treated in the relaxation-time approximation allowing for different relaxation times for quarks and gluons.   Baryon number conservation is enforced in the quark and anti-quark components of the fluid, but overall parton number non-conservation is allowed in the system.  The resulting equations are solved numerically in the (0+1)-dimensional boost-invariant case at zero and finite baryon density.
\end{abstract}



\maketitle 

\section{Introduction}
\label{sect:intro}

The dynamics of soft matter produced in ultra-relativistic heavy-ion collisions is well-described by relativistic viscous hydrodynamics \cite{Israel:1979wp,Muronga:2003ta,Baier:2006um,Romatschke:2007mq,Dusling:2007gi,Luzum:2008cw,Song:2008hj,Denicol:2010tr,Schenke:2011tv,Shen:2011eg,Bozek:2011wa,Niemi:2011ix,Bozek:2012qs}. However, due to the large gradients in the early stages of the collisions, viscous corrections to the ideal energy momentum tensor become quite substantial and the system becomes highly anisotropic in the momentum space, with the transverse pressure, $P_\perp$, typically much larger than the longitudinal pressure, $P_\parallel$. Large momentum-space anisotropies pose a problem for 2nd-order viscous hydrodynamics, since it relies on a linearization around an isotropic background. Large linear corrections generate unphysical results such as negative particle pressures, negative one-particle distribution functions, etc.~\cite{Martinez:2009mf}. This has motivated the development of reorganizations of viscous hydrodynamics in which one incorporates the possibility of large momentum-space anisotropies at leading order. The framework developed has been dubbed {\it anisotropic hydrodynamics} \cite{Florkowski:2010cf,Martinez:2010sc,Ryblewski:2010bs,Martinez:2010sd,Ryblewski:2011aq,Martinez:2012tu,Ryblewski:2012rr}. We note that anisotropic systems have been studied recently within the AdS/CFT correspondance framework \cite{Mateos:2011ix,Mateos:2011tv,Chernicoff:2012iq,Chernicoff:2012gu}.

In this paper we generalize the anisotropic hydrodynamics treatment to describe the dynamics of coupled quark and gluon fluids.  Our approach closely follows the formulation presented in Refs.~\cite{Martinez:2010sc,Martinez:2010sd,Martinez:2012tu} which is based on kinetic theory with the collisional kernel treated in the relaxation time approximation.  In this work the quark, antiquark, and gluon relaxation times are assumed to be constant.  We additionally assume that the system is boost invariant and homogeneous in the transverse directions which results in (0+1)-dimensional dynamics.   As a result we can take the quark, antiquark, and gluon distribution functions to be of spheroidal form \cite{Romatschke:2003ms}; however, we allow for different dynamical anisotropy parameters in the quark and gluon sectors.

We begin by setting up the problem in terms of kinetic theory and then obtain the necessary dynamical equations by taking the zeroth and first moments of the Boltzmann equation.  We then require baryon number conservation in the quark/anti-quark sector.  After doing this we find that the resulting system of equations is underdetermined.  In order to resolve this problem we reduce the number of independent variables by requiring that the quark and anti-quark anisotropy parameters are equal.

The structure of the paper is as follows: In Sec.~\ref{sect:kin} we introduce the kinetic equations for quarks, antiquarks, and gluons in the relaxation time approximation. In Sec.~\ref{sect:zeromoment} we discuss the zeroth moments of the kinetic equations. The baryon number and energy-momentum conservation laws are analyzed in Secs. \ref{sect:baryoncons} and \ref{sect:ene-mom-con}, respectively. In Sec.~\ref{sect:zero-bar-den} the case of zero baryon density is discussed in greater detail. The results of our numerical calculations are presented in Sec.~\ref{sect:resultsb0} and \ref{sect:resultsfb}, for the cases with zero and non-zero baryon density, respectively. We conclude in Sec.~\ref{sect:concl}.

\section{Kinetic equations}
\label{sect:kin}

In this work, we consider kinetic equations for the phase space distribution functions of quarks and antiquarks, $Q^\pm (x,p)$, and gluons, $G(x,p)$, given in the relaxation time approximation

\begin{equation}
 p^{\mu }\partial_{\mu } Q^\pm (x,p)= 
- p^\mu U_\mu \frac{Q^\pm(x,p) - Q^\pm_{\rm eq}(x,p)}{\tauq},  
\label{kineq0}
\end{equation}
\begin{equation}
p^{\mu }\partial_{\mu } G(x,p) = 
- p^\mu U_\mu \frac{G(x,p) - G_{\rm eq}(x,p)}{\taug}.  
\label{kineg0}
\end{equation} 
Here $\tauq$ and $\taug$ are the relaxation times for quarks and gluons, respectively, and $U^\mu$ is the flow four-vector
\begin{equation}
U^\mu = \gamma (1, v_x, v_y, v_z), \quad \gamma=(1-v^2)^{-1}.
\end{equation}
In this paper we will consider a system that is transversely homogenous and undergoing only boost-invariant longitudinal expansion.   In this case, the form of $U^\mu$ is fixed by the symmetry, and hence, below we set \mbox{$v_x=v_y=0$} and $v_z=z/t$.

Herein we will assume that the distribution functions are given by the covariant version of the Romatschke-Strickland distribution with $f_{\rm iso}$ given by a Boltzmann distribution \cite{Romatschke:2003ms}
\begin{eqnarray}
Q^\pm(x,p) &=& \exp\left(\frac{\pm \lambda - \sqrt{(p\cdot U)^2 + \xi_q^\pm (p\cdot V)^2}}{\Lambda} \right),
\nonumber \\
G(x,p) &=& \exp\left(-\frac{\sqrt{(p\cdot U)^2 + \xi_g (p\cdot V)^2}}{\Lambda} \right),
\label{RSform}
\end{eqnarray}
where the transverse-momentum scale $\Lambda$ is common for all of the distributions, while the anisotropy parameters $\xi^{\pm}_q$ and $\xi_g$ are allowed to be different. The parameter $\lambda$ plays a role of the baryon chemical potential.

The corresponding equilibrium distributions $Q^\pm_{\rm eq}(x,p)$ and $G_{\rm eq}(x,p)$ are defined by the expressions
\begin{eqnarray}
Q^\pm_{\rm eq}(x,p) &=& \exp\left(\frac{\pm \mu - p\cdot U}{T}  \right),
\nonumber \\
G_{\rm eq}(x,p) &=& \exp\left(-\frac{p\cdot U}{T}  \right),
\label{eqforms}
\end{eqnarray}
where $T$ is the temperature of the background and $\mu$ is the baryon chemical potential of quarks. The values of $T$ and $\mu$ follow from the Landau matching conditions which originate from the energy-momentum and baryon number conservation laws. 

The four-vector $V^\mu$ appearing in (\ref{RSform}) defines the direction of the beam ($z$-axis)
\begin{equation}
V^\mu = \gamma_z (v_z, 0, 0, 1), \quad \gamma_z = (1-v_z^2)^{-1/2}.
\label{V}
\end{equation}
We note that the four-vectors $U^\mu$ and $V^\mu$ satisfy the normalization conditions
\begin{eqnarray}
U^2 = 1, \quad V^2 = -1, \quad U \cdot V = 0.
\label{UVnorm}
\end{eqnarray}
In the local rest frame of the fluid element, $U^\mu$ and $V^\mu$ have simple forms
\begin{eqnarray}
 U^\mu = (1,0,0,0), \quad V^\mu = (0,0,0,1). 
 \label{UVLRF}
\end{eqnarray}
For the (0+1)-dimensional boost-invariant expansion considered in this paper, we may use
\begin{eqnarray}
 U^\mu &=& (\cosh\eta,0,0,\sinh\eta), \nonumber \\
 V^\mu &=& (\sinh\eta,0,0,\cosh\eta),
 \label{UVbinv}
\end{eqnarray}
where $\eta$ is the space-time rapidity
\begin{eqnarray}
\eta = \frac{1}{2} \ln \frac{t+z}{t-z}. \label{eta} 
\end{eqnarray}

\section{Zeroth moments of the kinetic equations}
\label{sect:zeromoment}

Integrating Eqs.~(\ref{kineq0}) and (\ref{kineg0}) over three-momentum and including the internal degrees of freedom we obtain the three equations
\begin{eqnarray}
\partial_{\mu } N_{q}^{\pm \, \mu} &=& 
 \frac{U_\mu \left( N_{q, \rm eq}^{\pm \, \mu} - N_{q}^{\pm \, \mu} \right)}{\tauq}, 
\label{zmq} \\
 \partial_{\mu } N_{g}^{\mu} &=& 
 \frac{U_\mu \left( N_{g, \rm eq}^{ \mu} - N_{g}^{\mu} \right)}{\taug},
\label{zmg}
\end{eqnarray}
where $N_{q}^{\pm \, \mu}$ and $N_{g}^{\mu}$ are particle currents
\begin{eqnarray}
N_{q}^{\pm \, \mu} &=& n_{q}^{\pm} U^\mu, \quad N_{g}^{\mu} = n_g U^\mu, 
\label{Nmuq} \\
N_{q, \rm eq}^{\pm \, \mu} &=& n_{q, \rm eq}^{\pm} U^\mu, \quad N_{g, \rm eq}^{\mu} = n_{g, \rm eq} U^\mu.
\label{Nmug}
\end{eqnarray}
This leads to the equations for the densities
\begin{eqnarray}
\partial_{\mu } (n_{q}^{\pm} U^{\mu}) &=& 
 \frac{n_{q, \rm eq}^{\pm} - n_{q}^{\pm}}{\tauq}, 
\label{denq} \\
 \partial_{\mu } (n_{g} U^{\mu} ) &=& 
 \frac{n_{g, \rm eq} - n_{g}}{\taug},
\label{deng}
\end{eqnarray}
where
\begin{eqnarray}
n_{q}^{\pm} &=& \frac{g_q}{\pi^2} \frac{e^{\pm \lambda/\Lambda} \Lambda^3}{\sqrt{1+\xi_q^\pm}}, 
\quad n_{q, \rm eq}^{\pm} = \frac{g_q}{\pi^2} e^{\pm \mu/T} T^3, 
\label{nq} \\
n_{g} &=& \frac{g_g}{\pi^2} \frac{ \Lambda^3}{\sqrt{1+\xi_g}}, 
\quad n_{g, \rm eq} = \frac{g_g}{\pi^2} T^3.
\label{ng}
\end{eqnarray}
The factors $g_q$ and $g_g$ in Eqs.~(\ref{nq}) and (\ref{ng}) account for internal degrees of freedom connected with spin and color. One can notice that they cancel out in (\ref{denq}) and (\ref{deng}). Substituting (\ref{nq}) and (\ref{ng}) into (\ref{denq}) and (\ref{deng}), respectively, we find the first dynamical equations, namely
\begin{eqnarray}
&&\frac{3}{\Lambda} \frac{d\Lambda}{d\tau}-\frac{1}{2(1+\xi_q^\pm)}\frac{d\xi_q^\pm}{d\tau} 
\pm \frac{d}{d\tau} \frac{\lambda}{\Lambda}+ \frac{1}{\tau}
\nonumber \\
&& \hspace{0.5cm}
=  \frac{1}{\tauq} \left(
\frac{e^{\pm (\mu/T-\lambda/\Lambda)} \, T^3}{\Lambda^3}\,  \sqrt{1+\xi_q^\pm}  - 1 \right), 
\label{EQ1q}
\end{eqnarray}
for quarks, and
\begin{eqnarray}
&&\frac{3}{\Lambda} \frac{d\Lambda}{d\tau}-\frac{1}{2(1+\xi_g)}\frac{d\xi_g}{d\tau} 
+ \frac{1}{\tau} 
\nonumber \\
&& \hspace{2cm} 
= \frac{1}{\taug} \left(
\frac{T^3}{\Lambda^3}\,  \sqrt{1+\xi_g}  - 1 \right),
\label{EQ1g}
\end{eqnarray}
for gluons.

Equations (\ref{EQ1q}) and (\ref{EQ1g}) are three equations for seven unknown functions of the proper time: $\xi_q^+$, $\xi_q^-$, $\xi_g$, $\Lambda$, $\lambda$, $T$, and $\mu$. The remaining four equations will be obtained from the baryon and energy-momentum conservation laws discussed in the next two Sections.

\section{Baryon number conservation}
\label{sect:baryoncons}

The difference of Eqs. (\ref{denq}) for quarks and antiquarks, multiplied by a factor of 1/3, yields 
\begin{equation}
\partial_\mu (b U^\mu) = \frac{b_{\rm eq}-b}{\tauq},
\label{b1}
\end{equation}
where $b$ is the quark baryon density
\begin{eqnarray}
b &=& \frac{1}{3} \left( n^+_q - n^-_q\right) \nonumber \\
&=& \frac{g_q \Lambda^3}{3\pi^2} \left(
\frac{e^{\lambda/\Lambda}}{\sqrt{1+\xi_q^+}} 
- \frac{e^{-\lambda/\Lambda}}{\sqrt{1+\xi_q^-}} \right),
\label{b2}
\end{eqnarray}
and $b_{\rm eq}$ is the baryon density of the equilibrium background 
\begin{eqnarray}
b_{\rm eq} &=& \frac{1}{3} \left( n^+_{q, \rm eq} - n^-_{q, \rm eq}\right) \nonumber \\
&=& \frac{g_q T^3}{3\pi^2} \left( e^{\mu/T}-e^{-\mu/T} \right).
\label{b3}
\end{eqnarray}
By demanding that the baryon number is conserved, 
\begin{equation}
\partial_\mu(b U^\mu)=0 
\label{baryoncon}, 
\end{equation}
we obtain the Landau matching condition for the baryon density 
\begin{equation}
b = b_{\rm eq}.
\label{b4}
\end{equation}
Equation (\ref{b4}) implies that the right-hand-sides of Eqs.~(\ref{b2}) and (\ref{b3}) are equal. This results in a constraint connecting different parameters of our model. 

To proceed further, we note that for (0+1)-dimensional expansion, the conservation law (\ref{baryoncon}) has the Bjorken-type solution
\begin{equation}
b(\tau) = b_i \frac{\tau_i}{\tau},
\label{b5}
\end{equation}
which may be identified with the right-hand-side of (\ref{b2}). Here $\tau_i$ is the initial proper time and $b_i$ is the corresponding initial baryon number density.

One may check that the use of the baryon number conservation (\ref{baryoncon}) implies that Eqs.~(\ref{denq})  are no longer independent. This suggests that the anisotropic distributions (\ref{RSform}) contain too many parameters to be determined within our scheme. In order to have the same numbers of independent equations and free parameters, from now on, we assume that the quark and antiquark anisotropies are the same, 
\begin{equation}
\xi_q^+ = \xi_q^- = \xi_q.
\label{xiq}
\end{equation}

As a consequence, we find
\begin{eqnarray}
\frac{\lambda}{\Lambda} &=& \sinh^{-1} (D) =
 \ln \left[D + \sqrt{1+D^2} \, \right],
\label{lambda}
\end{eqnarray}
where, in order to simplify our notation, we have introduced the quantity
\begin{equation}
D =  \frac{3  \pi^2 b \sqrt{1+\xi_q}}{2 g_q \Lambda^3}.
\label{D}
\end{equation}
Similarly, we find
\begin{equation}
\frac{\mu}{T} = \sinh^{-1} \left( \frac{D}{\kappa_q} \right) =
 \ln \left[\frac{D}{\kappa_q} + \sqrt{1+\frac{D^2}{\kappa_q^2}} \, \right],
\label{mu}
\end{equation}
where
\begin{equation}
\kappa_q =  \frac{T^3 \sqrt{1+\xi_q}}{\Lambda^3}.
\label{kappaq}
\end{equation}
In an analogous way we define
\begin{equation}
\kappa_g =  \frac{T^3 \sqrt{1+\xi_g}}{\Lambda^3}.
\label{kappag}
\end{equation}

Using Eqs.~(\ref{lambda}) and (\ref{mu}) in the ``quark'' equation in (\ref{EQ1q}) one finds
\begin{eqnarray}
&& \left( 1 - \frac{D}{\sqrt{1+D^2}} \right) \left[ \frac{3}{\Lambda} \frac{d\Lambda}{d\tau}
-\frac{1}{2(1+\xi_q)}\frac{d\xi_q}{d\tau} 
+ \frac{1}{\tau} \right]
\label{EQ1qb1} \\
&& = \left( 1 - \frac{D}{\sqrt{1+D^2}} \right) 
\frac{\left[ \kappa_q \sqrt{1+D^2} \sqrt{1+\frac{D^2}{\kappa_q^2}} 
- D^2 - 1 \right]}{\tauq} .
\nonumber
\end{eqnarray}
Using Eqs.~(\ref{lambda}) and (\ref{mu}) in the ``antiquark'' equation in (\ref{EQ1q}) one finds the same equation except that the factor $1 - D/\sqrt{1+D^2}$ is replaced by $1 + D/\sqrt{1+D^2}$. Since the factors $1 \pm D/\sqrt{1+D^2}$ may be canceled, Eqs.~(\ref{EQ1q}) can be reduced to a single equation 
\begin{eqnarray}
&&  \frac{3}{\Lambda} \frac{d\Lambda}{d\tau}
-\frac{1}{2(1+\xi_q)}\frac{d\xi_q}{d\tau} 
+ \frac{1}{\tau} 
\label{EQ1qb2} \\
&& \hspace{1cm} = 
\frac{1}{\tauq} \left[ \kappa_q \sqrt{1+D^2} \sqrt{1+\frac{D^2}{\kappa_q^2}} 
- D^2 - 1 \right].
\nonumber
\end{eqnarray}
%

\section{Energy-momentum conservation law}
\label{sect:ene-mom-con}

In order to introduce the energy-momentum conservation law for our system, we first sum over the quark, antiquark, and gluon degrees of freedom using Eqs.~(\ref{kineq0}) and (\ref{kineg0}), then we multiply this sum by $p^\nu$, and finally integrate over three-momentum. In this way, we obtain
\begin{eqnarray}
\partial_\mu T^{\mu \nu} = 
\frac{U_\mu \left(T_{q, \rm eq}^{\mu \nu}-T_q^{\mu \nu}\right)}{\tauq} +
\frac{U_\mu \left(T_{g, \rm eq}^{\mu \nu}-T_g^{\mu \nu}\right)}{\taug},
\label{enmomcon0}
\end{eqnarray}
where 
\begin{eqnarray}
T^{\mu \nu}=T_q^{\mu \nu}+T_g^{\mu \nu},
\label{Tmunuqpg}
\end{eqnarray}
and \cite{Florkowski:2010cf,Martinez:2012tu}
\begin{eqnarray}
T_i^{\mu \nu} &=& (\varepsilon_i+P_{i,\perp}) U^\mu U^\nu - P_{i, \perp} g^{\mu \nu} \nonumber \\
&& -(P_{i,\perp}-P_{i,\parallel}) V^\mu V^\nu \, ,
\label{Tmunu}
\end{eqnarray}
where the index $i$ stands for quarks ($i=q$) or gluons ($i=g$). In the similar way, we write
\begin{equation}
T_{i, \rm eq}^{\mu \nu} = (\varepsilon_{i, \rm eq}+P_{i, \rm eq}) U^\mu U^\nu - P_{i, \rm eq} g^{\mu \nu}.
\label{Tmunueq}
\end{equation}
The energy densities appearing in the formulas above have the form
\begin{eqnarray}
\varepsilon_q &=& \frac{6 g_q \Lambda^4}{\pi^2}  \cosh(\lambda/\Lambda) {\cal R}(\xi_q),
\nonumber \\
\varepsilon_g &=& \frac{3 g_g \Lambda^4}{\pi^2} {\cal R}(\xi_g),
\label{epsilon} \\
\varepsilon &=& \varepsilon_q + \varepsilon_g. \nonumber
\end{eqnarray}
The function ${\cal R}(\xi)$ appearing in (\ref{epsilon}) has the form \cite{Martinez:2010sc}
\begin{equation}
{\cal R}(\xi) = \frac{1}{2(1+\xi)} 
\left[1+ \frac{ (1+\xi) \arctan \sqrt{\xi}} {\sqrt{\xi} } \right].
\end{equation}
Similar expressions can be given for the transverse and longitudinal pressures. In our case, the important quantity is the (longitudinal) enthalpy, $\varepsilon + P_\parallel$, which can be written in the compact form
\begin{eqnarray}
\varepsilon_q + P_{q, \parallel} &=& -\frac{12 g_q \Lambda^4}{\pi^2}  \cosh(\lambda/\Lambda) (1+\xi_q) {\cal R}^\prime (\xi_q), \nonumber \\
\varepsilon_g + P_{g, \parallel} &=& -\frac{6 g_g \Lambda^4}{\pi^2} (1+\xi_g){\cal R}^\prime(\xi_g),
\label{enthalpy} \\
\varepsilon + P_\parallel &=& \varepsilon_q + P_{q, \parallel} + \varepsilon_g + P_{g, \parallel}.  \nonumber
\end{eqnarray}
Here, the prime denotes a derivative of the function ${\cal R}(\xi)$ with respect to its argument. The transverse pressure may be calculated from Eqs. (\ref{epsilon}) and (\ref{enthalpy}). The trace of the energy-momentum tensor is zero for massless particles. The latter condition gives \mbox{$\varepsilon = 2 P_\perp + P_\parallel$}.

In analogy to Eq.~(\ref{epsilon}), we obtain the formula for the energy density of the equilibrium background
\begin{eqnarray}
\varepsilon_{q, \rm eq} &=& \frac{6 g_q T^4}{\pi^2} \cosh(\mu/T), \nonumber \\
\varepsilon_{g, \rm eq} &=& \frac{3 g_g T^4}{\pi^2}, \label{epsiloneq} \\
\varepsilon_{\rm eq} &=& 
\varepsilon_{q, \rm eq} + \varepsilon_{g, \rm eq}. \nonumber
\end{eqnarray}
In order to conserve energy and momentum, the right-hand-side of (\ref{enmomcon0}) should vanish. Hence, we obtain the Landau matching condition for the energy density
\begin{equation}
\varepsilon_{q}  + 
\frac{\tauq}{\taug}\, \varepsilon_{g}  = 
\varepsilon_{q, \rm eq} +
\frac{\tauq}{\taug}\, \varepsilon_{g, \rm eq}.
\label{LM2}
\end{equation}
This leads us to another constraint 
\begin{eqnarray}
T^4 = \Lambda^4 \, \frac{ 2 \cosh(\lambda/\Lambda) {\cal R}(\xi_q)  + {\bar r} \, {\cal R}(\xi_g) }{2 \cosh(\mu/T)  + {\bar r}},
\label{T}
\end{eqnarray}
where we have introduced the coefficient 
\begin{eqnarray}
{\bar r} = \frac{\tauq}{\taug}\,
r 
\label{barr}
\end{eqnarray}
and
\begin{equation}
r = \frac{g_g}{g_q}.
\end{equation}
Clearly, when we compare the quark and gluon contributions to the thermalization processes, the value of $\bar r$ gives the relative weight of the gluon contribution.  The gluons are more important if their relaxation time is shorter. The Casimir scaling suggests that the mean free paths and, consequently, the relaxation times satisfy the relation
\begin{eqnarray}
 \frac{\tauq}{\taug} 
 = \frac{C_A}{C_F} = \frac{9}{4}.
 \label{caf}
\end{eqnarray}
Moreover, in our numerical calculations we use the values \mbox{$g_q = 2 \cdot 2 \cdot 3 = 12$} and $g_g = 2 \cdot 8 = 16$, hence $r = 4/3$.

Equation (\ref{T}) can be written in the equivalent form as
\begin{eqnarray}
\frac{T^4}{\Lambda^4} =  \, \frac{ 2 \sqrt{1+D^2} \, {\cal R}(\xi_q)  + {\bar r} \, {\cal R}(\xi_g) }{2 \sqrt{1+ (3\pi^2 b/2 g_q T^3)^2}  + {\bar r}}.
\label{TL}
\end{eqnarray}
For a given set of the values of $\tau$, $\xi_q(\tau)$, $\xi_g(\tau)$, and $\Lambda(\tau)$, Eq.~(\ref{TL}) may be used to calculate the temperature of the equilibrium background, $T(\tau)$.

If the condition (\ref{LM2}) is satisfied, the left-hand-side of Eq.~(\ref{enmomcon0}) yields the conservation law for energy and momentum
\begin{equation}
\partial_\mu T^{\mu \nu} = 0.
\label{enmomcon}
\end{equation}
In our (0+1)-dimensional case, Eq.~(\ref{enmomcon}) is reduced to the expression
\begin{equation}
\frac{d\varepsilon}{d\tau} = -\frac{\varepsilon+P_\parallel}{\tau}.
\label{enmomcon01}
\end{equation}
Substituting Eqs.~(\ref{epsilon}) and (\ref{enthalpy}) in (\ref{enmomcon01}) one finds
\begin{eqnarray}
&&\frac{d}{d\tau} \left[ \Lambda^4 \left( 2 \sqrt{1+D^2} \,
{\cal R}(\xi_q)  + r {\cal R}(\xi_g) \right) \right] \label{eneq} \\
&& = \frac{2 \Lambda^4}{\tau} \left[ 2 \sqrt{1+D^2} \, (1+\xi_q) {\cal R}^\prime (\xi_q)  + r (1+\xi_g) {\cal R}^\prime (\xi_g) \vphantom{e^{\lambda/\Lambda}} \right]. \nonumber
\end{eqnarray}
Since we have eliminated the dependence on $\lambda$ and $\mu$, Eqs.~(\ref{EQ1g}), (\ref{EQ1qb2}),  (\ref{TL}), and (\ref{eneq}) represent four equations for four unknown functions: $\xi_q$, $\xi_g$, $T$, and $\Lambda$. 

\section{Zero baryon density}
\label{sect:zero-bar-den}

We consider first the special case of vanishing baryon number. The condition $b=0$ used in Eqs. (\ref{lambda}) and (\ref{mu}) leads to the conclusions
\begin{equation}
\lambda = 0, \quad \mu = 0.
\label{lambda0mu0}
\end{equation}
Hence, we are left with the following four equations for four unknown functions ($\xi_q$, $\xi_g$, $T$, and $\Lambda$)
\begin{eqnarray}
\hspace{-3mm} \frac{3}{\Lambda} \frac{d\Lambda}{d\tau}-\frac{1}{2(1+\xi_q)}\frac{d\xi_q}{d\tau} 
+ \frac{1}{\tau} =  \frac{\kappa_q  - 1}{\tauq} ,
\label{EQ1q0}
\end{eqnarray}
\begin{eqnarray}
\hspace{-3mm} \frac{3}{\Lambda} \frac{d\Lambda}{d\tau}-\frac{1}{2(1+\xi_g)}\frac{d\xi_g}{d\tau} 
+ \frac{1}{\tau}  =  \frac{\kappa_g - 1}{\taug} ,
\label{EQ1g0}
\end{eqnarray}
\begin{eqnarray}
\frac{T^4}{\Lambda^4} =  \frac{2 {\cal R}(\xi_q) + {\bar r} {\cal R}(\xi_g) }{2 + {\bar r}},
\label{T0}
\end{eqnarray}
\begin{eqnarray}
\label{EQecons}
&&\frac{d}{d\tau} \left[ \Lambda^4 \left( 
2 {\cal R}(\xi_q) + r {\cal R}(\xi_g) \right) \right] \label{eneq0} \nonumber  \\
&& = \frac{2 \Lambda^4}{\tau} \left[ 2 (1+\xi_q) {\cal R}^\prime (\xi_q) 
+ r (1+\xi_g) {\cal R}^\prime (\xi_g) \right].
\end{eqnarray}

The above system of equations can be reduced to two equations, if we calculate the ratio $T/\Lambda$ from Eq.~(\ref{T0}) and the derivative $\dot{\Lambda} = d\Lambda/d\tau$ from Eq. (\ref{eneq0}) and substitute these two quantities into Eqs.~(\ref{EQ1q0}) and (\ref{EQ1g0}). In this way we obtain,
\begin{eqnarray}
\frac{d\xi_q}{d\tau} &=& 2(1+\xi_q) \left[\frac{1}{\tau} + \frac{p_q(\xi_q,\xi_g)}{\tauq \, \Delta(\xi_q,\xi_g)} \right]
\label{EQ1q1}
\end{eqnarray}
and
\begin{eqnarray}
\frac{d\xi_g}{d\tau} &=& 2(1+\xi_g) \left[\frac{1}{\tau} + 
\frac{p_g(\xi_q,\xi_g)}{\taug \, \Delta(\xi_q,\xi_g)} \right],
\label{EQ1g1}
\end{eqnarray}
where
\begin{eqnarray}
\Delta(\xi_q,\xi_g) &=& \frac{\pi^2 (P_\perp-P_\parallel) }{3 g_q \Lambda^4} 
= 4 {\cal R}(\xi_q) + 2 r {\cal R}(\xi_g)   \nonumber \\
&& \hspace{-10mm}  + \, 6 \, (1+\xi_q) {\cal R}^\prime(\xi_q) + 3 r (1+\xi_g) {\cal R}^\prime (\xi_g),
\label{Delta}
\end{eqnarray}
\begin{eqnarray}
p_q(\xi_q,\xi_g) &=&  (1-\kappa_q) \left[4 {\cal R}(\xi_q) + 2 r {\cal R}(\xi_g) \right]  \label{pq} \\
&&\hspace{-10mm} + \, \frac{3 r}{C_F} \left[C_F (1-\kappa_q) - C_A (1-\kappa_g)\right] (1+\xi_g) {\cal R}^\prime (\xi_g),
\nonumber
\end{eqnarray}
\begin{eqnarray}
p_g(\xi_q,\xi_g) &=&  (1-\kappa_g) \left[4 {\cal R}(\xi_q) + 2 r {\cal R}(\xi_g) \right]   \label{pg} \\
&& \hspace{-10mm}+ \,  \frac{6}{C_A} \left[C_A (1-\kappa_g)- C_F (1-\kappa_q) \right]  (1+\xi_q) {\cal R}^\prime (\xi_q).
\nonumber
\end{eqnarray}

One may check that
\begin{eqnarray}
\frac{p_q}{\tauq}-\frac{p_g}{\taug} &=&  \Delta \left(
 \frac{\kappa_g-1}{\taug}
-\frac{\kappa_q-1}{\tauq}
\right) \label{pqminuspg} \\
&=&  \frac{\Delta}{C_A \taug}
\left[C_A (\kappa_g-1) - C_F (\kappa_q-1)
\right].
\nonumber
\end{eqnarray}
%

\subsection{Zero baryon density:  late-time behavior}
\label{subsec:latetime}

\begin{figure}[t]
\includegraphics[angle=0,width=0.45\textwidth]{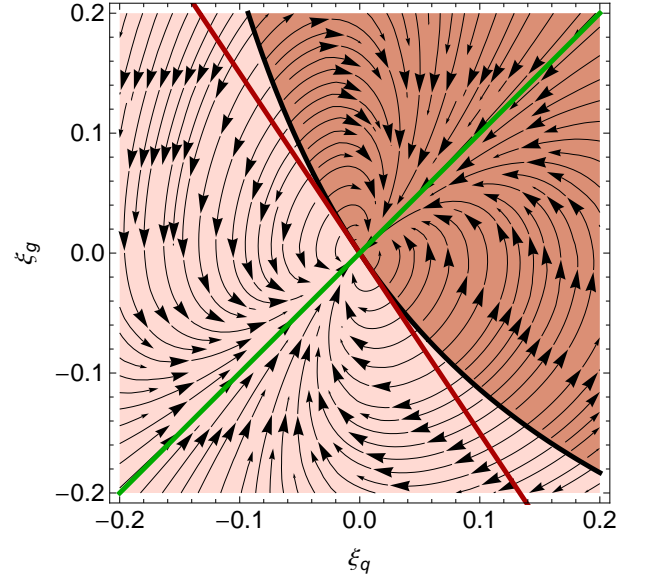}
\caption{(Color online) Phase space plot of the late time dynamics of the coupled dynamical equations for
$\xi_q$ and $\xi_g$.  The black line shows the line $\Delta(\xi_q,\xi_g)=0$. The red line shows the line
$\xi_g = -(2/r)\xi_q$.  The green line lindicates the line $\xi_g=\xi_q$.  The darker shaded region corresponds
to $\Delta(\xi_q,\xi_g)<0$ and the lighter shaded region to $\Delta(\xi_q,\xi_g)>0$.  For this plot we have
used the Casimir scaling $\tauq/\taug = 9/4$ corresponding to $N_c = 3$.}
\label{fig:fp}
\end{figure}

Before proceeding to numerical solutions we will investigate the late-time behavior of the dynamical equations for the quark and 
gluon fluids.  In order to gain a qualitative understanding of the dynamics one expects at late times, in 
Fig.~\ref{fig:fp} we present a vector plot of the right hand sides of Eqs.~(\ref{EQ1q1}) and (\ref{EQ1g1}) when
$\tau=10^{12}$ fm/c.  The 
vectors shown indicate the magnitude and phase-space direction of the time derivatives of $\xi_q$ and $\xi_g$.  The black 
line shows the line $\Delta(\xi_q,\xi_g)=0$ with $\Delta(\xi_q,\xi_g)$ defined in Eq.~(\ref{Delta}). The red line 
shows the line $\xi_g = -(2/r)\xi_q$.  The green line lindicates 
the line $\xi_g=\xi_q$.  The darker shaded region corresponds to $\Delta(\xi_q,\xi_g)>0$ and the lighter 
shaded region to $\Delta(\xi_q,\xi_g)<0$.   

As can be seen from this figure, the line $\Delta(\xi_q,\xi_g)=0$ is a repulsive line, with points to the ``right'' 
of this line flowing to the right and points to the ``left'' of this line flowing to the left.  In this way the system 
dynamically avoids the 
line of singularities corresponding to $\Delta(\xi_q,\xi_g)=0$ (assuming that the system is not initialized with 
exactly  $\Delta(\xi_q,\xi_g)=0$).\footnote{There is a subtle exception to this when $\tauq=\taug$ and $\xi_q = \xi_g$.
In this case, it is possible to cross the line corresponding to $\Delta=0$; however, there is no singularity
in this case since the right hand sides of Eqs.~(\ref{EQ1q1}) and (\ref{EQ1g1}) are finite at $\xi_q = \xi_g = 0$.}

Since at late times both anisotropy parameters tend towards zero, one can perform small anisotropy expansions
in order to determine the precise nature of the late-time behavior of the dynamical equations.
Based on empirical analysis of numerical solutions to Eqs.~(\ref{EQ1q0}),  (\ref{EQ1g0}), and 
(\ref{EQecons}) there are two cases to be considered:  (I) $\lim_{\tau \rightarrow \infty} \xi_q 
\neq \lim_{\tau \rightarrow \infty} \xi_g$ with $\lim_{\tau \rightarrow \infty} \xi_q/\xi_g < 0$ 
and (II) $\lim_{\tau \rightarrow \infty} \xi_q = \lim_{\tau \rightarrow \infty} \xi_g = \xi > 0$.
We will analyze the late-time solutions of the system in these two cases separately because the
analytic expansions necessary turn out to be quite different.

\subsubsection{Case I}

In this case we begin by taking the difference of Eqs.~(\ref{EQ1q0}) and (\ref{EQ1g0}) to 
obtain\footnote{We could have also taken the difference of (\ref{EQ1q1}) over $1+\xi_q$ and
(\ref{EQ1g1}) over $1+\xi_g$ and obtained the same equation.}
\begin{equation}
\frac{\dot{\xi}_q}{1+\xi_q} - \frac{\dot{\xi}_g}{1+\xi_g}
 =  \frac{2\left[\taug(\kappa_g-1)-\tauq(\kappa_q-1)\right] }{\tauq\,\taug} \, ,
\label{EQ1-latetime}
\end{equation}
where a dot indicates a derivative with respect to proper time.
Since both $\xi_q$ and $\xi_g$ are assumed to be small at late times, we can Taylor expand the equations
in order to understand the late-time behavior.
Using Eqs.~(\ref{kappaq}), (\ref{kappag}), and (\ref{T0}) we can expand to linear order
in $\xi_q$ and $\xi_g$ to obtain
\begin{equation}
\dot{\xi}_q - \dot{\xi}_g = - \gamma (\xi_q - \xi_g) \, ,
\label{deltadotm}
\end{equation}
where $\gamma \equiv \alpha \taug^{-1}$ with
\begin{equation}
\alpha \equiv   \frac{2+r}{2+\bar{r}} \, .
\end{equation}
We can use this to solve for $\xi_q$ in terms of $\xi_g$
\begin{equation}
\xi_q = \xi_g + A e^{-\gamma \tau} \, ,
\label{linxiq}
\end{equation}
where $A$ is an undetermined constant.
To proceed, we use Eq.~(\ref{EQecons}) to eliminate $\dot{\Lambda}$ in Eqs.~(\ref{EQ1q0}) 
and (\ref{EQ1g0}) and expand to linear order in $\xi_q$ and $\xi_g$ to obtain
\begin{eqnarray}
\dot{\xi}_g - \dot{\xi}_q &\!=\!& 
\left(\frac{4}{15\tau} + \gamma \right) \! \xi_q
+ \left(\frac{2 r}{15\tau} - \gamma \right) \! \xi_g \, , 
\nonumber \\
\dot{\xi}_q - \dot{\xi}_g &\!=\!& 
\left(\frac{4}{15\tau} + \gamma \right) \! \xi_g
+ \left(\frac{8}{15 r \tau} - \gamma \right) \! \xi_q \, .
\end{eqnarray}
Adding these two equations we find $\xi_q = - r \xi_g/2$.
Using (\ref{linxiq}) we can then find the explicit forms for $\xi_q$ and $\xi_g$
\begin{eqnarray}
\lim_{\tau \rightarrow \infty} \xi_q &=& -\frac{1}{2} B r e^{-\gamma\tau} \, , 
\nonumber \\
\lim_{\tau \rightarrow \infty} \xi_g &=& B e^{-\gamma\tau} \, , 
\end{eqnarray}
where $B$ is a constant.  
Plugging these solutions back into Eq.~(\ref{EQecons}) and taking the large-$\tau$ limit one obtains
\begin{equation}
\lim_{\tau \rightarrow \infty} \Lambda(\tau) = \frac{C}{\tau^{1/3}} \, ,
\end{equation}
where $C$ is an undetermined constant.
These results describe
the late-time behavior of the numerical solutions in Case I very well.  As we can see from 
the expressions above, in this case one finds that at late times the system will, at some point, 
rapidly approach isotropy.

We note that in order to assess the stability of the solution obtained at leading order 
one can extend the solution next-to-leading order in the anisotropy parameters.  To do this 
we expand $\xi_q$ and $\xi_g$ as
\begin{eqnarray}
\xi_q = -\frac{1}{2} B r e^{-\gamma \tau} + \delta \xi_q \, , \nonumber \\
\xi_g = B e^{-\gamma\tau} + \delta \xi_g \, ,
\end{eqnarray}
treating the first terms as ${\cal O}(\xi_{q,g})$ and the second terms as ${\cal O}(\xi_{q,g}^2)$.
Inserting these and expanding the resulting equations to quadratic order, one obtains the following
solution for $\delta \xi_g$
\begin{equation}
\delta \xi_g = D e^{-\gamma\tau} + E \tau e^{-2\gamma\tau} \, ,
\end{equation}
where $D$ and $E$ are undetermined constants.  Since the first term can be absorbed into the leading-order 
late-time behavior, we can set $D=0$.  One can also solve for $\delta \xi_q$ in the same manner to 
obtain
\begin{equation}
\delta \xi_q = E \tau e^{-2\gamma\tau} \, .
\end{equation}
Since both perturbations, $\delta \xi_g$ and $\delta \xi_q$, decrease exponentially, the leading order
solution $\xi_q = -r \xi_g/2$ is a stable fixed point of the late-time dynamics in Case I.

\begin{figure}[t]
\begin{center}
\subfigure{\includegraphics[angle=0,width=0.45\textwidth]{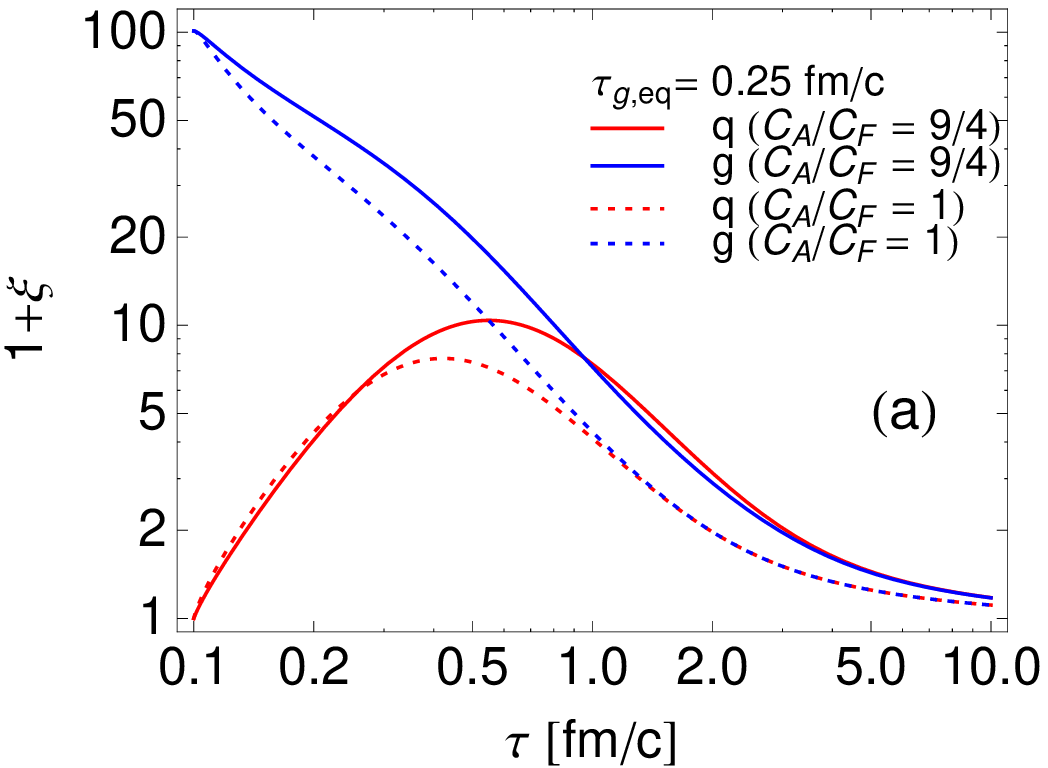}}\\
\subfigure{\includegraphics[angle=0,width=0.45\textwidth]{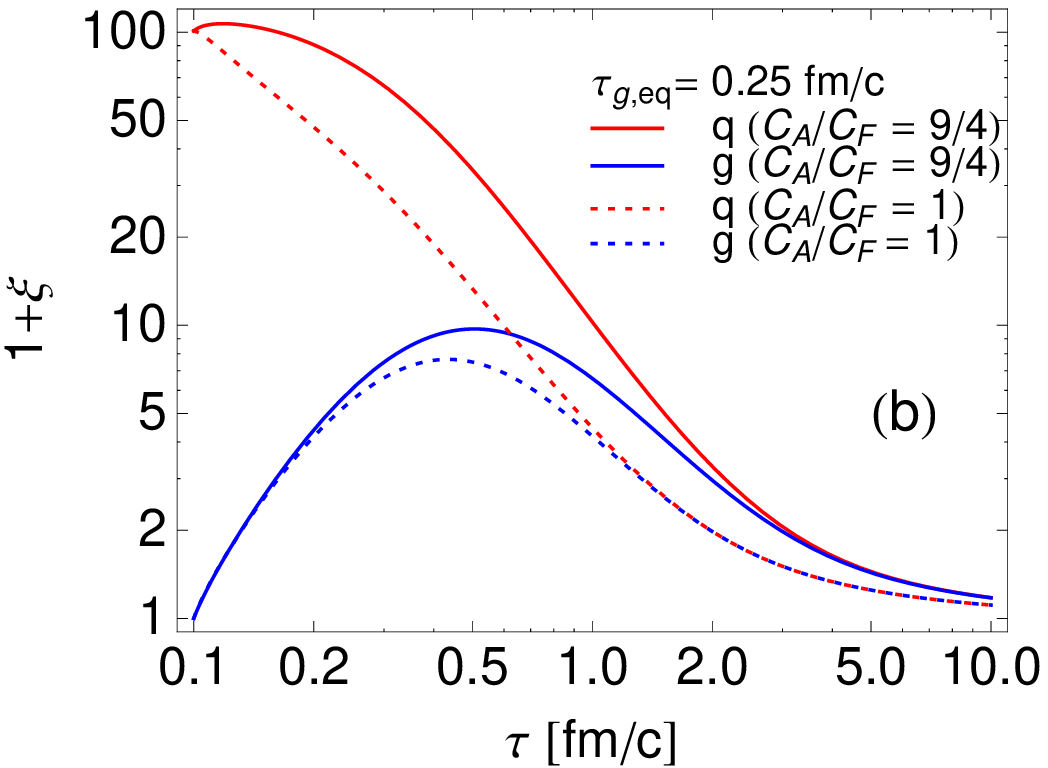}}
\end{center}
\caption{(Color online) Time dependence of the anisotropy parameters of gluons (solid blue lines) and quarks (solid red lines) for the initial conditions $\xi_g = 100$ and $\xi_q=0$ (a) and $\xi_q = 100$ and $\xi_g=0$ (b).}
\label{fig:f_100_0001}
\end{figure}

\begin{figure}[t]
\begin{center}
\subfigure{\includegraphics[angle=0,width=0.45\textwidth]{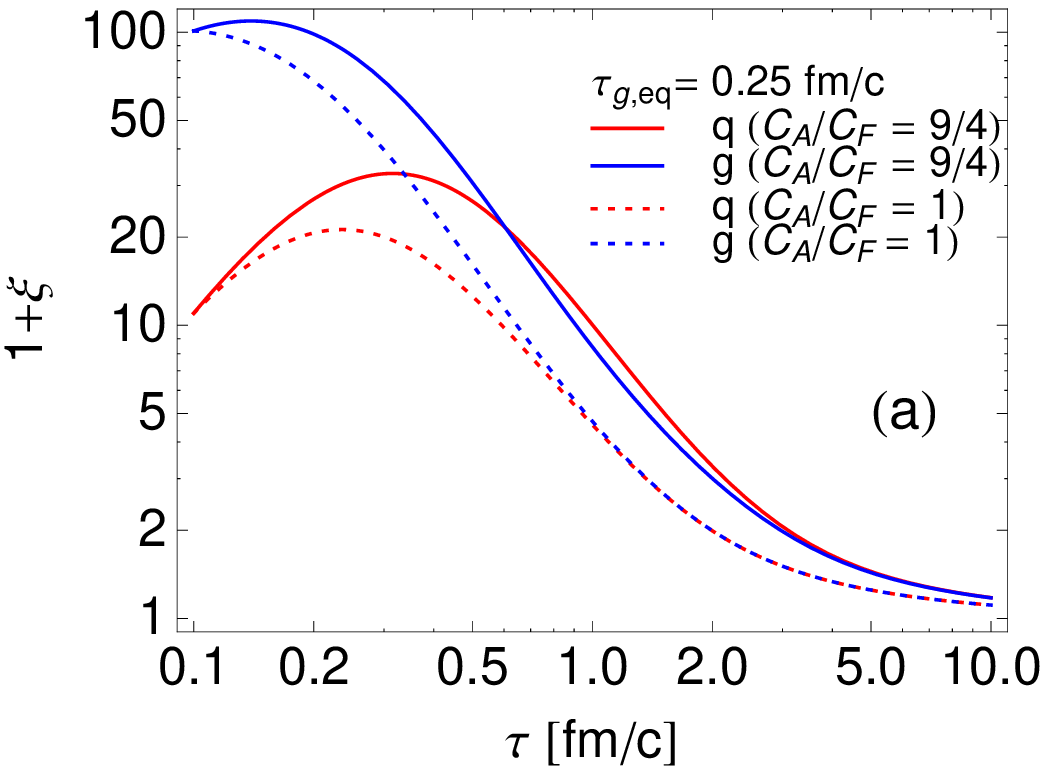}}\\
\subfigure{\includegraphics[angle=0,width=0.45\textwidth]{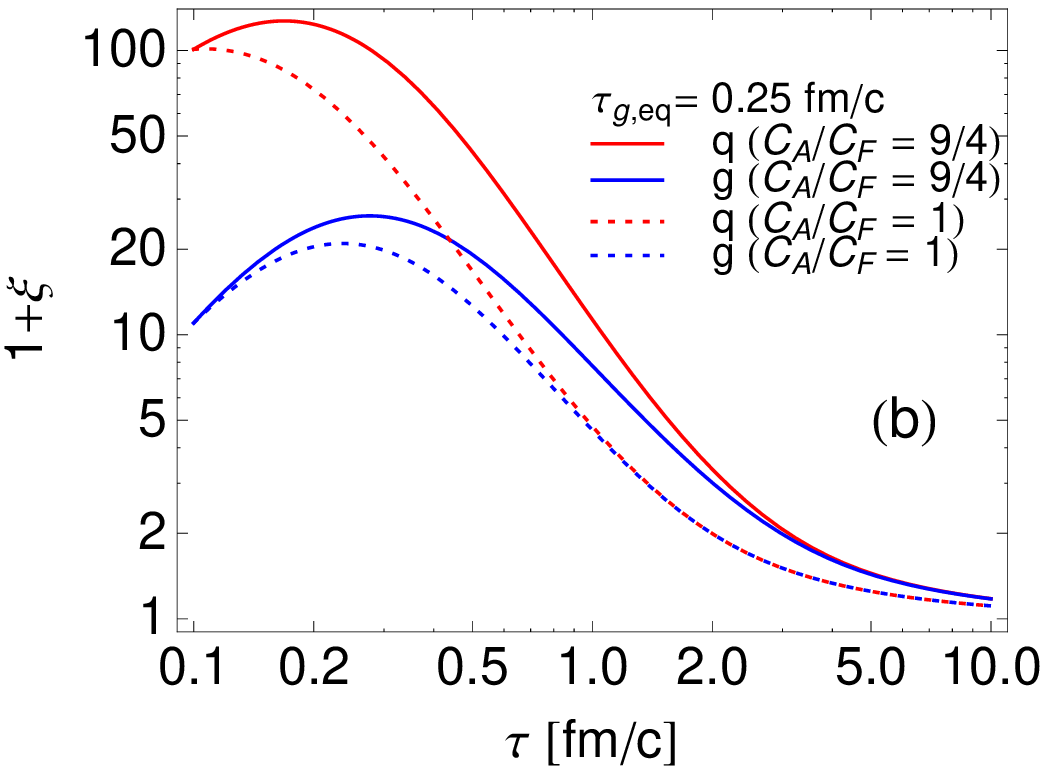}}
\end{center}
\caption{(Color online) The same as Fig.~\ref{fig:f_100_0001} but for the initial conditions $\xi^i_g = 100$ and $\xi^i_q=10$ (a) and $\xi^i_q = 100$ and $\xi^i_g=10$ (b).}
\label{fig:f_100_10}
\end{figure}

\subsubsection{Case II}

In this case one finds power-law decays at late times, so we instead search for
power law solutions of the form
\begin{eqnarray}
\xi_q &=& A_1 \tau^{-1} + A_2 \tau^{-2} + A_3 \tau^{-3} + {\cal O}(\tau^{-4})  \, , \\
\xi_g &=& B_1 \tau^{-1} + B_2 \tau^{-2} + B_3 \tau^{-3} + {\cal O}(\tau^{-4}) \, .
\end{eqnarray}
As in Case I we use Eq.~(\ref{EQecons}) to eliminate $\dot{\Lambda}$ in Eqs.~(\ref{EQ1q0}) 
and (\ref{EQ1g0}) and then plug in the ansatz above.  We then make a large-$\tau$ expansion
of the equations.  Requiring the ${\cal O}(\tau^{-1})$ to vanish results in $A_1 = B_1$.  One
then proceeds order by order requiring terms of order $\tau^{-n}$ to vanish, which
uniquely fixes the coefficients at the preceding order.  The resulting solutions through ${\cal O}(\tau^{-2})$ are
\begin{equation}
A_1 = B_1 = 4 \alpha \tauq \, ,
\end{equation}
and
\begin{eqnarray}
A_2 &=& \frac{8\tauq}{315} \biggl[ 42 (\alpha-1)\taug  + (84 + 521 \alpha)\alpha \tauq \biggr] , \nonumber \\
B_2 &=& \frac{8\tauq}{315} \biggl[ 42 (2\alpha-1)\taug  + (42 + 521 \alpha)\alpha \tauq \biggr] . \nonumber \\
\end{eqnarray}
From this one can derive a compact expression for the difference of $\xi$'s at large times
\begin{equation}
\xi_q - \xi_g = \frac{16}{15} \alpha \tauq (\tauq-\taug) \tau^{-2} \, .
\end{equation}
Note that in the case $\tauq=\taug=\tau_{\rm eq}$, this correction vanishes identically.  In fact, one 
finds in this special case that all power law corrections vanish and instead the difference
$\xi_q  - \xi_g$ goes to zero exponentially at late times.

Inserting the general solutions obtained above for $\xi_q$ and $\xi_g$ into Eq.~(\ref{EQecons}) and taking the 
large-$\tau$ limit one obtains
\begin{equation}
\lim_{\tau \rightarrow \infty} \Lambda(\tau) = \frac{C}{\tau^{1/3}} \, ,
\end{equation}
where $C$ is an undetermined constant.

We have compared the asymptotic solutions specified above with direct numerical solution at late times in Case II, and
in all cases checked we found excellent agreement. 

\section{Results for the zero baryon density}
\label{sect:resultsb0}

In this Section we discuss our results obtained in the case of vanishing baryon number density, $b=0$. The solutions of Eqs.~(\ref{EQ1q1}) and (\ref{EQ1g1}) are presented for a few possible options for the initial conditions. 

In Fig.~\ref{fig:f_100_0001} we show the time dependence of the anisotropy parameters of gluons (solid blue lines) and quarks (solid red lines). The initial time for anisotropic hydrodynamic evolution is $\tau_i$ = 0.1 fm/c. The relaxation times in the kinetic equations are constant:  \mbox{$\tau_{g, \rm  eq}$ = 0.25 fm/c} and \mbox{$\tau_{q, \rm  eq} = (9/4) \, \tau_{g, \rm  eq}$}. In Fig.~\ref{fig:f_100_0001} (a), the initial values are $\xi^i_g = 100$ and $\xi^i_q=0$, while in Fig.~\ref{fig:f_100_0001} (b)  the initial values are reversed, $\xi^i_g = 0$ and $\xi^i_q=100$. The dotted lines indicate the time evolution of the anisotropy parameters in the case where the two relaxation times are equal, \mbox{$\tau_{q, \rm  eq} = \tau_{g, \rm  eq}$ =  0.25 fm/c}. In the latter case the average relaxation time is shorter, which is reflected in a faster decay of the gluon and quark anisotropy in the cases (a) and (b), respectively.


\begin{figure}[t]
\begin{center}
\subfigure{\includegraphics[angle=0,width=0.45\textwidth]{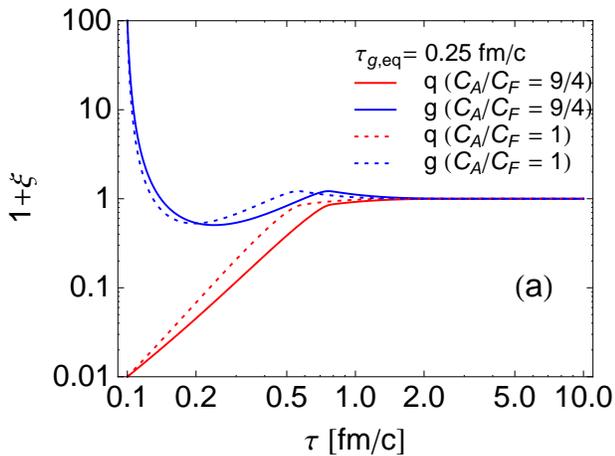}}\\
\subfigure{\includegraphics[angle=0,width=0.45\textwidth]{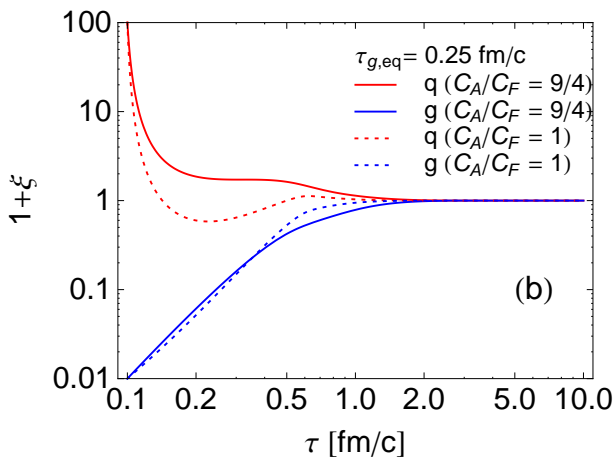}}
\end{center}
\caption{(Color online) The time evolution of the quark and gluon anisotropies for the initially mixed prolate-oblate configurations: $\xi^i_g = 100$ and $\xi^i_q=-0.99$ (a) and $\xi^i_q = 100$ and $\xi^i_g=-0.99$ (b).}
\label{fig:f_100_-99}
\end{figure}

\begin{figure}[t]
\begin{center}
\subfigure{\includegraphics[angle=0,width=0.45\textwidth]{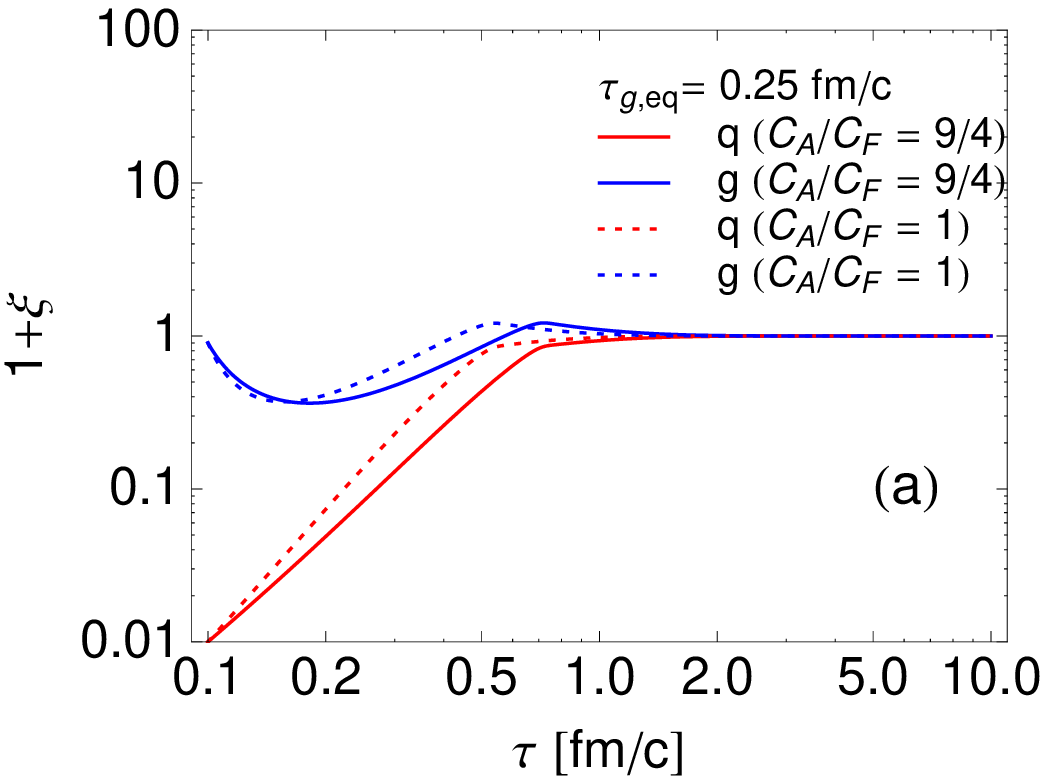}}\\
\subfigure{\includegraphics[angle=0,width=0.45\textwidth]{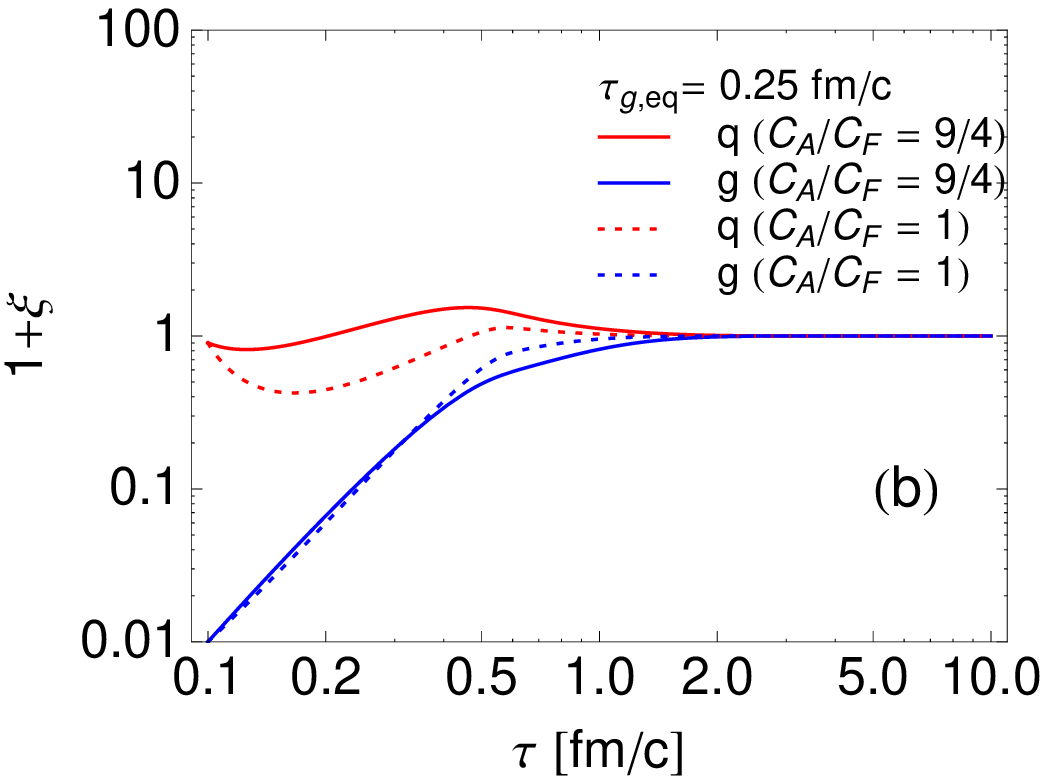}}
\end{center}
\caption{(Color online) The time evolution of the quark and gluon anisotropies for the initial prolate configurations: $\xi^i_g = -0.1$ and $\xi^i_q=-0.99$ (a) and $\xi^i_q = -0.1$ and $\xi^i_g=-0.99$ (b).}
\label{fig:f_-0.1_-0.99}
\end{figure}

\begin{figure}[t]
\begin{center}
\subfigure{\includegraphics[angle=0,width=0.45\textwidth]{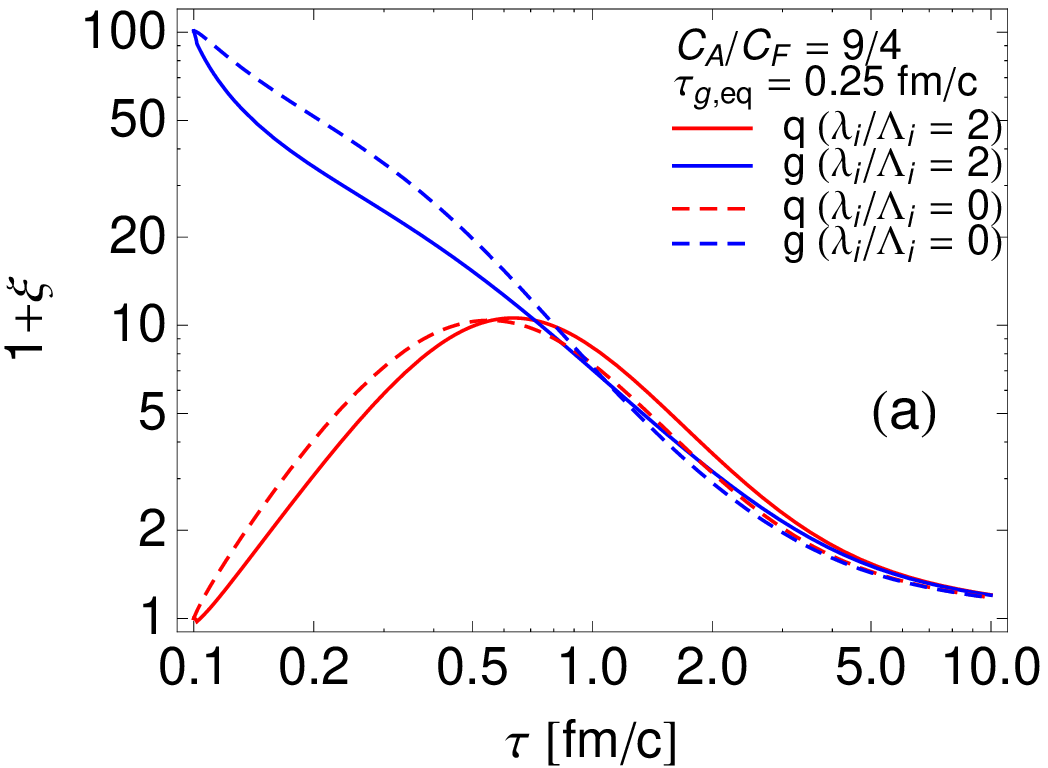}}\\
\subfigure{\includegraphics[angle=0,width=0.45\textwidth]{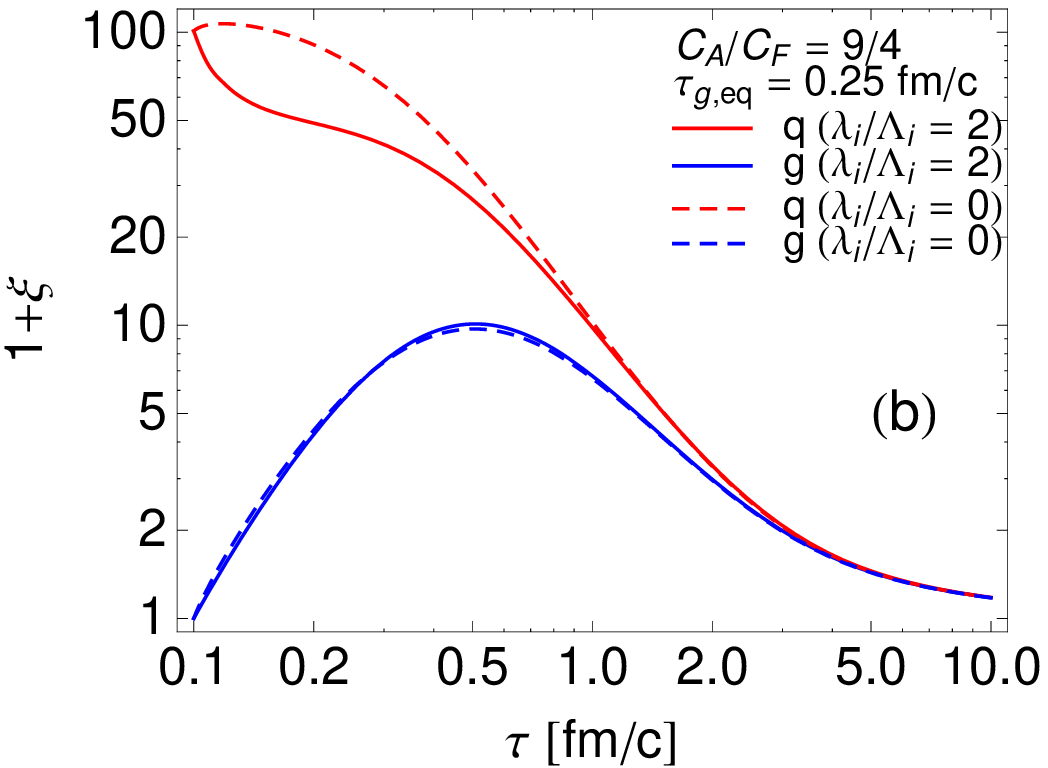}}
\end{center}
\caption{(Color online) The same as Fig.~\ref{fig:f_100_0001} but with the inclusion of the finite baryon number density corresponding to $\lambda_i/\Lambda_i=2$. The dashed lines describe the time dependence of the anisotropy parameters in the case $b=0$.}
\label{fig:f_100_0001_B}
\end{figure}

\begin{figure}[t]
\begin{center}
\subfigure{\includegraphics[angle=0,width=0.45\textwidth]{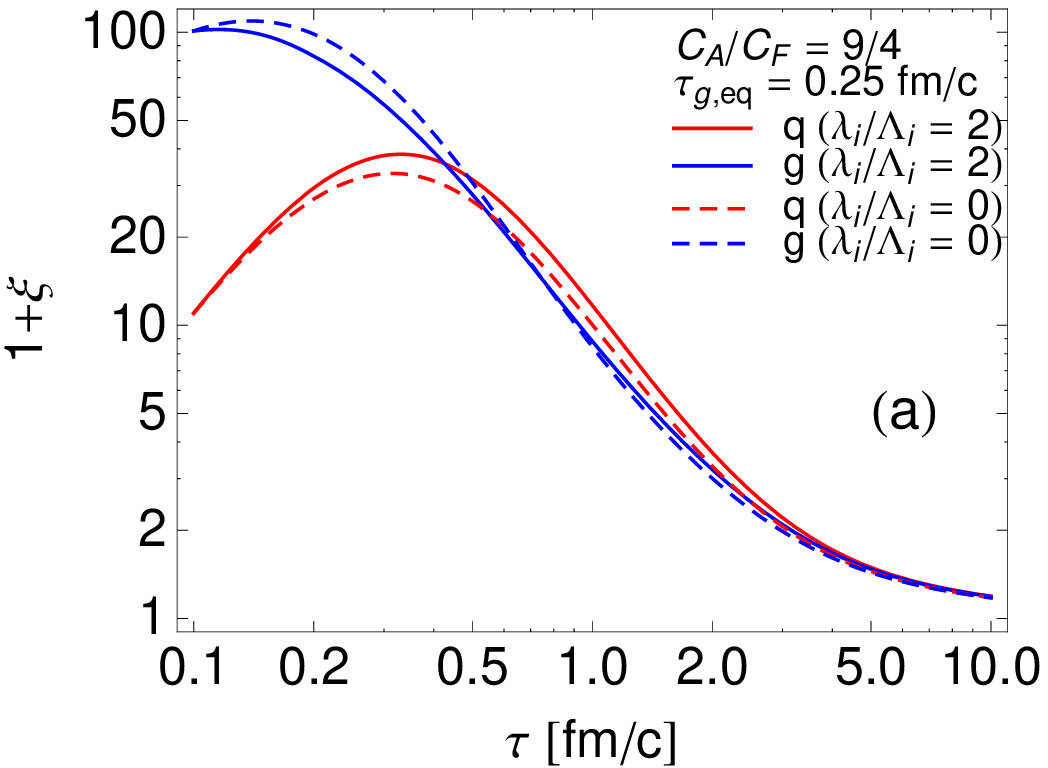}}\\
\subfigure{\includegraphics[angle=0,width=0.45\textwidth]{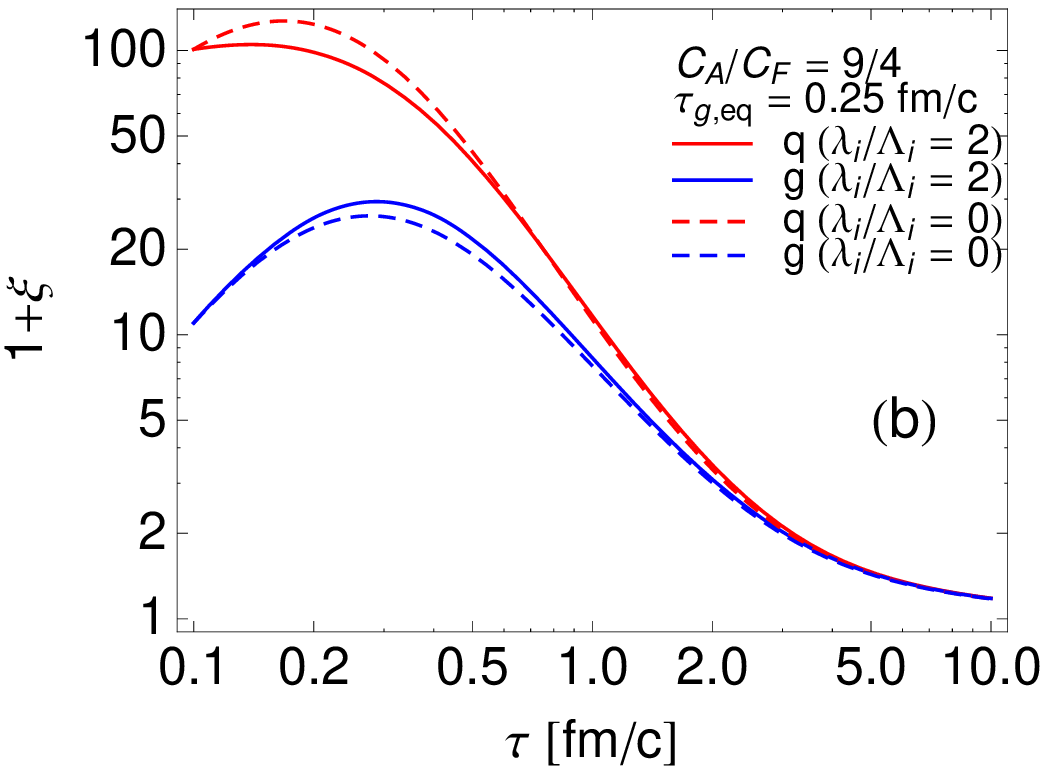}}
\end{center}
\caption{(Color online) The same as Fig.~\ref{fig:f_100_10} but with the inclusion of the finite baryon number density corresponding to $\lambda_i/\Lambda_i=2$. The dashed lines describe the time dependence of the anisotropy parameters in the case $b=0$.}
\label{fig:f_100_10_B}
\end{figure}


Interestingly,  Fig.~\ref{fig:f_100_0001} shows that the anisotropy of a less anisotropic subsystem is increased through its interaction with a more anisotropic subsystem.  Only later, the two systems equilibrate. This behavior is naturally expected, on the grounds of the overall tendency of the whole system to reach equilibrium. 

In Fig.~\ref{fig:f_100_10} we show again the time dependence of the anisotropy parameters of gluons (solid blue lines) and quarks (solid red lines). In this case the initial conditions are $\xi^i_g = 100$ and $\xi^i_q=10$ in (a) and $\xi^i_q = 100$ and $\xi^i_g=10$ in (b). Since now both the gluon and quark initial anisotropies are large, the thermalization effects are not so strong as to decrease the anisotropies at the beginning of the evolution. The very early dynamics is dominated by $1/\tau$ terms in (\ref{EQ1q1}) and (\ref{EQ1g1}) which are responsible for free-streaming effects leading to an increase of the anisotropy. 

The systems described in Figs.~\ref{fig:f_100_0001} and \ref{fig:f_100_10} correspond to systems which are initially {\it oblate} (the first anisotropy parameter is very large and positive, while the second parameter is positive or equals zero). For comparison, it is interesting to consider also the cases where one or two anisotropy parameters are negative. Such cases correspond to mixed {\it oblate-prolate} or {\it prolate} configurations. 

In Fig.~\ref{fig:f_100_-99} we show the results obtained for the initial conditions $\xi^i_g = 100$ and $\xi^i_q=-0.99$ (a) and $\xi^i_q = 100$ and $\xi^i_g=-0.99$ (b). In this case, we observe a very fast equilibration of matter. After about 0.5~fm/c, which corresponds to the evolution time $\tau \approx 2\tau_{g, \rm  eq}$, the scaling solution $\xi_q = -(r/2) \xi_g$ is reached with an exponential decay of both anisotropies -- see our earlier discussion of the Case I. Similar scaling behavior is observed in the cases where the two initial anisotropies are negative. This is shown in Fig.~\ref{fig:f_-0.1_-0.99}. We note that the systems which are initially oblate have the time asymptotics described by the Case II.

\section{Results for the finite baryon density}
\label{sect:resultsfb}

The equations valid for $b=0$ may be also used to a good approximation in the case of small baryon number density, defined by the conditions
\begin{equation}
\frac{b}{T^3} \ll 1, \quad \frac{\mu}{T} \ll 1,
\end{equation}
which, due to the Landau matching (\ref{b4}), imply also
\begin{equation}
\frac{b \, \sqrt{1+\xi_q}}{\Lambda^3} \ll 1, \quad \frac{\lambda}{\Lambda} \ll 1.
\end{equation}
This is so, because all corrections start with quadratic terms in $b$. In practice, the results obtained with the initial ratio $\lambda_i/\Lambda_i \sim 0.5$ are still very well approximated by the results obtained in the limit $b=0$. Therefore, to observe noticeable effects of baryon number conservation on the anisotropy evolution, we have to consider large initial values of the ratio $\lambda/\Lambda$.

In Fig.~\ref{fig:f_100_0001_B} we present our numerical results obtained for the initial ratio $\lambda_i/\Lambda_i = 2$. We show the time dependence of the anisotropy parameters of gluons (solid blue lines) and quarks (solid red lines), with the same initial values as those used in Fig.~\ref{fig:f_100_0001}, i.e., $\xi^i_g = 100$ and $\xi^i_q=0$ in (a), and $\xi^i_g = 0$ and $\xi^i_q=100$ in (b). The dashed lines describe the time dependence of the anisotropy parameters in the case $b=0$ (with the same initial conditions for anisotropy parameters). Compared to the case with $b=0$, we observe larger differences between the parts (a) and (b). Clearly, the baryon number conservation affects more strongly the evolution of quarks than the evolution of gluons --- see the differences between the red and blue lines in the part (b). Figure~\ref{fig:f_100_10_B} shows analogous results obtained with $\lambda_i/\Lambda_i = 2$, $\xi^i_g = 100$, and $\xi^i_q=10$ in (a), and $\xi^i_g = 10$ and $\xi^i_q=100$ in (b). 

We have also performed numerical calculations for the finite baryon density with the initial oblate-prolate and prolate configurations for $\lambda_i/\Lambda_i = 2$. Our results are similar to those obtained at zero baryon density and presented in Figs.~\ref{fig:f_100_-99} and \ref{fig:f_-0.1_-0.99}.

\section{Conclusions}
\label{sect:concl}

In this paper we have generalized the concept of anisotropic hydrodynamics to describe the dynamics of coupled quark and gluon fluids allowing for different dynamical anisotropy parameters for the quark and gluon fluids. Our approach has followed closely the formulation presented in \cite{Martinez:2010sc,Martinez:2010sd,Martinez:2012tu} which is based on kinetic theory with the collisional kernel treated in the relaxation time approximation except herein we allowed for the possibility of different relaxation times for the quark and gluon fluids. The resulting equations have been solved numerically in the (0+1)-dimensional boost-invariant case at zero and finite baryon density showing that both fluids tend towards isotropic fixed points at late times.  Additionally, for the case of zero baryon density we made a detailed analytic analysis of the late-time behavior of the solutions and obtained the asymptotic behavior to next-to-leading order.  

In our opinion, the results presented in this paper may be used to model the very early stages of ultra-relativistic heavy-ion collisions, where large pressure anisotropies are expected and the quark and gluon sectors may not have the same degree of momentum-space anisotropy.  As a next step it would be interesting to generalize the approach presented here to a non-boost-invariant case, where the rapidity profiles of the baryon density and the energy density are different. This may be done along the lines presented in \cite{Ryblewski:2010bs,Martinez:2010sd}.  It is also possible to relax the assumption that the isotropic distribution function is given by a Boltzmann distribution.  This will not affect the dynamical equations qualitatively, and should only require the generalization of some of the dynamical equations presented herein.  In addition, in the future one can also study the effect of including a relaxation time which is proportional to the inverse transverse momentum scale.  We leave these interesting problems for future work.

\begin{acknowledgments}

This work was supported by the Polish Ministry of Science and Higher Education under Grant No. N N202 263438 and the United States National Science Foundation under Grant No.~PHY-1068765.

\end{acknowledgments}

\bibliography{mixture}

\end{document}